\documentclass[aps,prd,reprint,preprintnumbers,showpacs,showkeys,superscriptaddress]{revtex4-1}
\usepackage{amsmath,amssymb,bm,mathrsfs,graphicx}
\usepackage[colorlinks=true,citecolor=blue,linkcolor=blue]{hyperref}

\begin{document}
\title{Nambu-Goldstone bosons with fractional-power dispersion relations}

\author{Haruki Watanabe}
\email{hwatanabe@berkeley.edu}
\affiliation{Department of Physics, University of California,
  Berkeley, California 94720, USA}

\author{Hitoshi Murayama}
\email{hitoshi@berkeley.edu, hitoshi.murayama@ipmu.jp}
\affiliation{Department of Physics, University of California,
  Berkeley, California 94720, USA} 
\affiliation{Theoretical Physics Group, Lawrence Berkeley National
  Laboratory, Berkeley, California 94720, USA} 
\affiliation{Kavli Institute for the Physics and Mathematics of the
  Universe (WPI), Todai Institutes for Advanced Study, University of Tokyo,
  Kashiwa 277-8583, Japan} 

\begin{abstract}
  We pin down the origin of a peculiar dispersion relation of domain
  wall fluctuation, the so-called ripplon, in a superfluid-superfluid
  interface.  A ripplon has a dispersion relation
  $\omega\propto k^{3/2}$ due to the nonlocality of the
  effective Lagrangian, which is mediated by gapless superfluid
  phonons in the bulk.  We point out the analogy to the longitudinal
  phonon in the two-dimensional Wigner crystal.
\end{abstract}

\preprint{IPMU14-0058,UCB-PTH14/05}
\maketitle

\section{Introduction}
The dispersion relation of Nambu-Goldstone bosons (NGBs) determines the low-energy property of systems with spontaneous symmetry breaking. It is directly connected to the temperature dependence of thermodynamic quantities, such as heat capacitance.  The softness of NGBs also sets the severeness of infrared divergence, from which one can determine the stability of the symmetry-breaking ground state~\cite{WatanabeMurayama5}.

The dispersion relation of NGBs is usually an integer power in the
long-wavelength limit; i.e., $\omega\propto
|k_i|^{n_i}$ with $n_i\in\mathbb{Z}$ for $\vec{k}=k_i\hat{x}_i$
($i=1,\ldots,d$; no sum over $i$), with an appropriate choice of the axes.  For example, phonons in an
ordinary crystal have a linear dispersion relation of $\omega\propto |\vec{k}|$ ($n_i=1$), and magnons in a
Heisenberg ferromagnet have a quadratic dispersion relation,
$\omega\propto |\vec{k}|^2$ ($n_i=2$).  In a helical magnet
with the spiral vector along the $z$ axis, the dispersion relation of
helimagnons takes an anisotropic form, $\omega\propto
\sqrt{k_z^2+C(k_x^2+k_y^2)^2}$~\cite{Radzihovsky}. Namely, it is
linear in the $z$ direction ($n_z=1$) and quadratic in $x,y$
directions ($n_x=n_y=2$).

We can actually ``prove" the integer power by assuming a {\it local}\/ effective Lagrangian. In Fourier space, the quadratic effective Lagrangian can be expressed as $(1/2)\pi^a(\vec{k},\omega)^*\mathcal{G}_{ab}^{-1}(\vec{k},\omega)\pi^b(\vec{k},\omega)$, where $\pi^a$'s are Nambu-Goldstone fields (see Ref.~\cite{WatanabeMurayama5} for more details).  For the wave vector $\vec{k}=k_i\hat{x}_i$, one finds
\begin{equation}
\mathcal{G}_{ab}^{-1}=\bar{g}_{ab}\omega^2+i\rho_{ab}\omega-g_{(i)ab}k_i^2-g_{(i)ab}'k_i^4+\cdots,\label{green}
\end{equation}
where $\rho$ is an antisymmetric matrix and $\bar{g}, g_{(i)},g_{(i)}'$ are symmetric matrices.
For example, when $\rho=0$, all NGBs have the usual linear dispersion
relation (type A).  When $\rho$ is nonzero and full-rank, we may
neglect the $\bar{g}_{ab}$ term, and all NGBs have a quadratic dispersion
relation (type B).   When some components of $g_{(i)ab}$ somehow
vanish, these NGBs may have softer dispersion relations, but they power is still an integer~\footnote{We may add a term $C_{(i)}'\omega k_i$ but the conclusion does not fundamentally change}. Crucially, Eq.~\eqref{green} does not have a term $C_{(i)ab}k_i$; as such a term would cause an instability toward a ground state with a nonzero $\vec{k}$. Even in this case, we can re-expand $\mathcal{G}_{ab}^{-1}$ from the momentum minimum $\vec{k}_0$ with no resulting linear term in $\vec{k}-\vec{k}_0$.

Nevertheless, it is known that the fluctuation of a certain domain wall
has a dispersion relation with a fractional power,
$\omega\propto k^{3/2}$~\cite{Mazets}.  Although there are
several analytical and numerical studies that support this weird
dispersion relation~\cite{Volovik,Takeuchi2,Bezett,Kobyakov,Takeuchi}, the
physically intuitive picture behind it remains unclear in
the existing literature.  In this paper, we pin down its origin as the
breakdown of the locality of the effective Lagrangian, resulting from
integrating out gapless modes in the bulk.  
In general, when the microscopic Lagrangian of a system is local, its effective Lagrangian obtained by integrating out only higher-energy modes  is still local.  However, when the system contains an interaction of two subsystems $S_{\text{A},\text{B}}$ of different dimensionality,  it may be useful to integrate out $S_{\text{B}}$ --- regardless of its gap --- to get an effective theory of $S_{\text{A}}$ alone.
A nonlocal effective Lagrangian in real space is equivalent to a non-analytic
dependence on momentum in Fourier space, which may lead to a
dispersion relation with a fractional power as we shall see below.  

\section{Effective Lagrangian of Ripplon}
We consider a domain wall in a two-component Bose-Einstein condensate.  When the
intercomponent repulsion is much stronger than the intracomponent
one, a spontaneous phase separation occurs and a domain wall is formed
between the components.  Since the domain wall spontaneously breaks the
translational symmetry in the direction perpendicular to the plane,
there should be a gapless Nambu-Goldstone mode corresponding to the
fluctuation (ripple) of the interface. Such a mode is referred to as a ``ripplon."  Here we rederive its dispersion relation by deriving an effective Lagrangian written solely in terms of the displacement
field $u$.

\subsection{Model}
Let us take a domain wall at $z=0$ in the equilibrium and describe its
fluctuation by a displacement field $u=u(\vec{x}_{\parallel},t)$ with
$\vec{x}_{\parallel}=(x,y)^{T}$. A superfluid $1$ (2) fills the space
$z>u$ $(z<u)$.  Each superfluid is described by the standard Lagrangian $\mathcal{P}_a(X_a)=\frac{X_a^2}{2g_a}$, with $X_a=\mu_a-\dot{\varphi}_a-(2m_a)^{-1}[(\vec{\nabla}_{\parallel}\varphi_a)^2+(\partial_z\varphi_a)^2]$ for $a=1,2$. (Here and hereafter, there is no sum over $a$ and $\hbar=1$.) Thus the phenomenological Lagrangian for this system should be given by~\cite{Takeuchi2,Takeuchi}
\begin{eqnarray}
\mathcal{L}_{\text{eff}}=\int_{u}^{\infty}\mathrm{d}z\,\mathcal{P}_1(X_1)+\int_{-\infty}^{u}\mathrm{d}z\,\mathcal{P}_2(X_2)-\sigma\mathcal{S},\label{start}
\end{eqnarray}
where $\mathcal{S}=\sqrt{1+(\vec{\nabla}_{\parallel}u)^2}$ is the area
element and $\sigma$ is the tension of the domain wall.  Here we take the space to be infinitely extended. $\varphi_a=0$ and $u=0$ characterize the ground state without a superflow.  We consider small fluctuations above it at zero temperature; hence there is no normal component of the fluid.

The variation of the Lagrangian~\eqref{start} with respect to $u$ gives Laplace's law, $(\mathcal{P}_1-\mathcal{P}_2)|_{z=u}=\sigma\vec{\nabla}_{\parallel}\cdot (\mathcal{S}^{-1}\vec{\nabla}_{\parallel}u)$, which relates the pressure difference across the surface to the surface tension.  The pressure balance at the ground state requires $\mu_1^2/2g_1=\mu_2^2/2g_2$.

Our goal is to derive an effective field theory of the surface fluctuation in terms of the displacement field $u(\vec{x}_{\parallel},t)$.  Our strategy is simply to integrate out bulk degrees of freedom $\varphi_{1,2}$.

\subsection{Equation of motion of superfluid phonons}
Let us first clarify the equation of motion for $\varphi_{1,2}$ fields
that describe Bogoliubov phonons in superfluids.  The variation of the
Lagrangian \eqref{start} with respect to $\varphi_a$ requires extra
care to the $u$ dependence of the integration domain. For example,
when we integrate by parts the time derivative in a combination
$\int_{-\infty}^{\infty}\mathrm{d}t\int_{-\infty}^u\mathrm{d}z\,f(z,t)\partial_t\delta\varphi(z,t)$,
$\partial_t$ may act on $u(\vec{x}_{\parallel},t)$, as well as on
$f(z,t)$ in the integrand. Therefore, the variation of the action with respect to $\varphi_a$ is
\begin{eqnarray}
\delta S_{\text{eff}} = 
(-1)^a\int\mathrm{d}t\mathrm{d}^2x_{\parallel}(j_a^z-\partial_\mu u j_a^\mu)\delta\varphi_a|_{z=u}\notag\\
-\int\mathrm{d}t\mathrm{d}^2x_{\parallel}\mathrm{d}z\left(\partial_\mu
  j_a^\mu\right)\delta\varphi_a.\label{variation}
\end{eqnarray}
The first line gives the boundary condition $j_a^z=\partial_\mu
u j_a^\mu$ at $z=u$.  Here, $j_a^\mu\equiv-\partial\mathcal{P}_a/\partial(\partial_\mu\varphi_a)$
is the conserved $\text{U}(1)$ current.  Using its linearized form,
the boundary condition is
\begin{equation}
\dot{u}=\left.\frac{\partial_z\varphi_a}{m_a}\right|_{z=0}.
\label{boundary}
\end{equation}
The physical meaning of this condition is clear: the domain wall and the superfluids have the same $z$ component of the velocity at the boundary.  This boundary condition is necessary for the conservation of U(1) charges, 
\begin{equation}
Q_1=\int\mathrm{d}^2x_{\parallel}\int_u^\infty\mathrm{d}z\,j_1^0,\quad
Q_2=\int\mathrm{d}^2x_{\parallel}\int_{-\infty}^u\mathrm{d}z\,j_1^0.\label{u1}
\end{equation}

The second line of Eq.~\eqref{variation} gives the equation of continuity $\partial_\mu j_a^\mu=0$. It can be linearized as
\begin{equation}
\ddot{\varphi}_a-v_a^2(\partial_{\parallel}^2+\partial_z^2)\varphi_a=0,\quad v_a^2\equiv\frac{\mu_a}{m_a}.
\end{equation}
We solve this  equation in the form
$\varphi(\vec{x}_{\parallel},z,t)=\varphi_a(\vec{k}_{\parallel},\omega)e^{i\vec{k}_{\parallel}\cdot\vec{x}_{\parallel}+(-1)^a\kappa_az-i\omega
  t}$, with
$\kappa_a=\sqrt{k_{\parallel}^2-(\omega/v_a)^2}$ and $k_{\parallel}=|\vec{k}_{\parallel}|$, assuming the fluctuation is localized on the domain wall. Combined with the condition in Eq.~\eqref{boundary}, we find an expression of $\varphi$ in terms of $u$:
\begin{eqnarray}
\varphi_a(\vec{k}_{\parallel},\omega)=(-1)^a\frac{m_a(-i\omega)}{\kappa_a}u(\vec{k}_{\parallel},\omega).\label{solphi}
\end{eqnarray}

\subsection{The effective domain wall Lagrangian}
To get the effective Lagrangian in terms of the displacement field $u$ only, we substitute the solution of $\varphi$ in Eq.~\eqref{solphi} back into the Lagrangian. This is equivalent to integrating out $\varphi_a$ at the tree level.  The bulk term $[\dot{\varphi}_a^2-(\vec{\nabla}_\parallel\varphi)^2-(\partial_z\varphi)^2]/2g_a$ does not contribute since this vanishes thanks to the equation of motion. The crucial contribution comes from
\begin{equation}
-\int_{u}^{\infty}\mathrm{d}z\,n_1\dot{\varphi}_1-\int_{-\infty}^{u}\mathrm{d}z\,n_2\dot{\varphi}_2
\end{equation}
in the Lagrangian \eqref{start}, where $n_a=\mu_a/g_a$ is the superfluid (number) density.  After integrating by parts, we obtain
\begin{equation}
(-n_1\varphi_1+n_2\varphi_2)|_{z=0}\,\dot{u}\label{coupling}
\end{equation}
at the linearized level.  This combination is quite intriguing since
it makes $\dot{u}$ and $(-n_1\varphi_1+n_2\varphi_2)|_{z=0}$
canonically conjugate to each other (at least when we neglect higher-order terms $\dot{\varphi}_a^2$). Such a conjugate relation usually
leads to noncommuting symmetry algebra, although all symmetries under
consideration [$\text{U}(1)_a$ and translations] are naively Abelian. We will further discuss this point
later.

Putting the pieces together, we find 
\begin{eqnarray}
\mathcal{L}_{\text{eff}}=\frac{1}{2}u^*\left(\frac{m_1n_1\omega^2}{\sqrt{k_{\parallel}^2-\frac{\omega^2}{v_1^2}}}+\frac{m_2n_2\omega^2}{\sqrt{k_{\parallel}^2-\frac{\omega^2}{v_2^2}}}-\sigma k_{\parallel}^2\right)u\label{effective}
\end{eqnarray}
in Fourier space~\footnote{Reference~\cite{Volovik} presented a very similar expression, which neglects the $\omega^2/v_a^2$ term in the denominator. This neglected term becomes important when we take gravity into account.}.  In the long-wavelength limit, the Lagrangian
correctly describes the known dispersion relation
$\omega^2=\sigma
k_{\parallel}^3/(m_1n_1+m_2n_2)$.  The
$\omega^2/v_a^2$ term in the denominator can be consistently neglected
for a dispersion relation $k_{\parallel}^n$ ($n>1$) in the long-wavelength limit.

The effective Lagrangian in real space reads
\begin{eqnarray}
L_{\text{eff}}&=&\frac{1}{2}\int\mathrm{d}^2x_{\parallel}\mathrm{d}^2y_{\parallel}\,\dot{u}(\vec{x}_{\parallel},t)\frac{m_1n_1+m_2n_2}{2\pi|\vec{x}_{\parallel}-\vec{y}_{\parallel}|}\dot{u}(\vec{y}_{\parallel},t)\nonumber\\
&&\quad\quad-\int\mathrm{d}^2x_{\parallel}\,\frac{1}{2}\sigma[\vec{\nabla}_{\parallel}u(\vec{x}_{\parallel},t)]^2\label{real}
\end{eqnarray}
to leading order in derivatives.  One can observe a Coulomb-type long-range coupling between the
velocity of the domain wall.  Hence, the effective Lagrangian is
nonlocal, invalidating our ``proof" of the integer power for a local
effective Lagrangian.  It may be an interesting future work to
examine the Coulomb-type long-range coupling from the dual picture
of superfluids.  This peculiar form of the time-derivative term may
also have other interesting physical consequences, e.g.,
response to an external force.

\subsection{Discussions}
\begin{figure}
\begin{center}
\includegraphics[width=0.8\columnwidth]{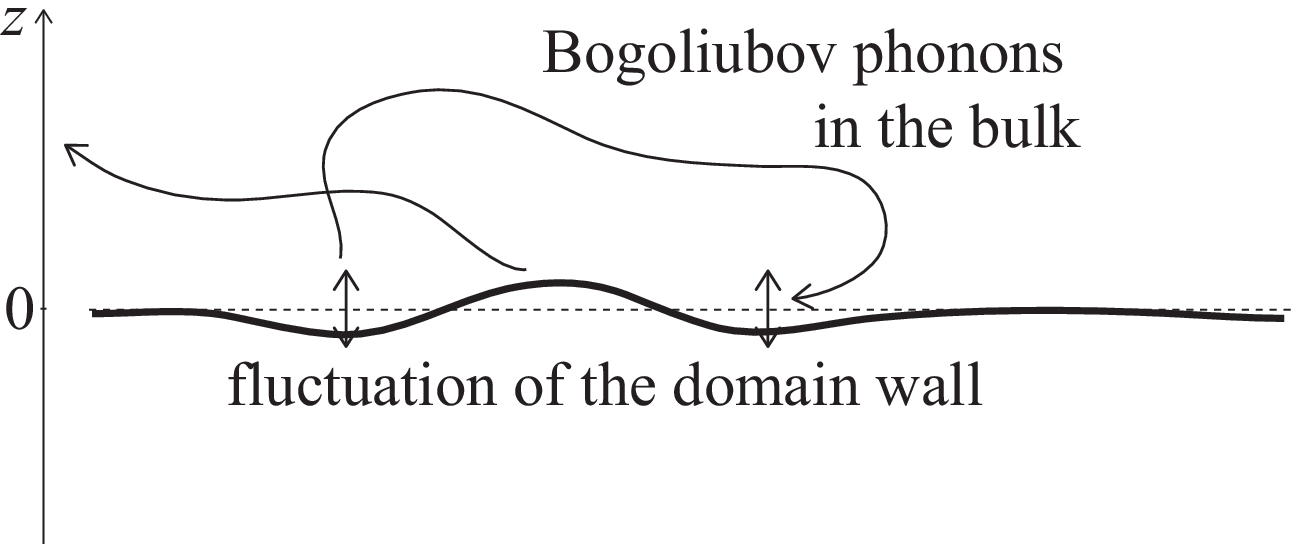}
\includegraphics[width=0.8\columnwidth]{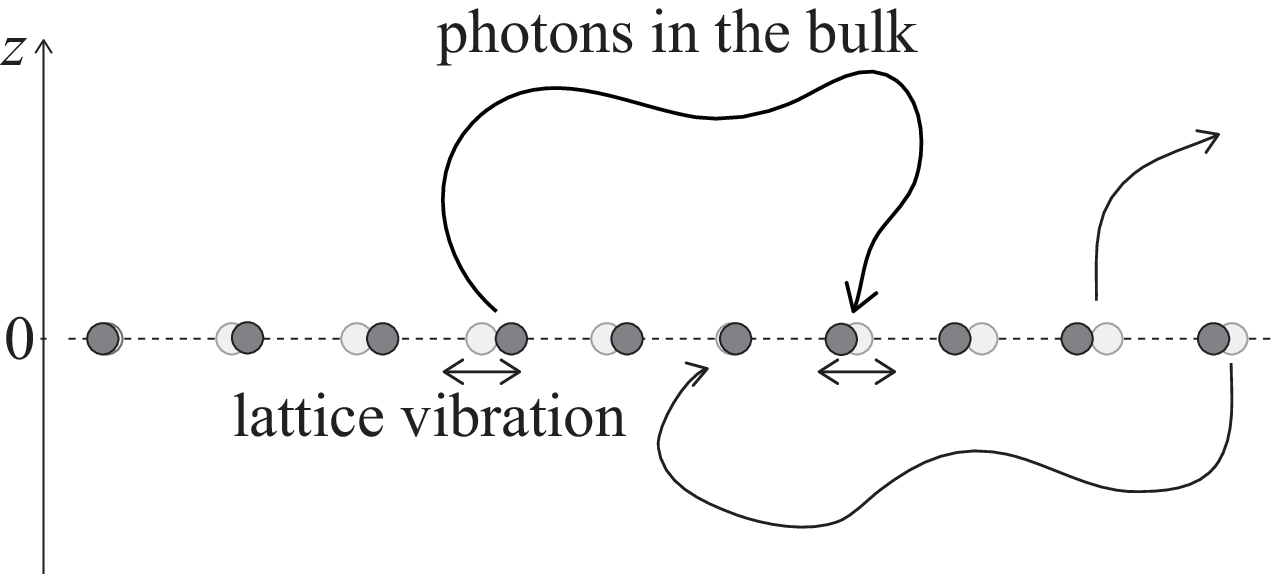}
\end{center}
\caption{(Top panel) Fluctuation of the domain wall (ripple) excites Bogoliubov phonons in the bulk, and the bulk excitation in turn affects the domain wall fluctuation, inducing an effective nonlocal interaction. (Bottom panel) The system of lattice vibration (phonons) of a Wigner crystal and gauge photons in the bulk is perfectly analogous to the ripple example.}
\label{fig}
\end{figure}

The nonlocality of the effective Lagrangian clearly originates from
integrating out gapless bulk modes.  When the bulk mode is gapless, a low-energy fluctuation of the domain wall excites bulk modes and, in turn, the bulk oscillation affects a different part of the domain wall, as illustrated in Fig.~\ref{fig}. This is the physical picture behind the nonlocal term mediated by gapless bulk modes.  Then the Fourier component of the nonlocal term $\int\mathrm{d}^dx\mathrm{d}^dy\,\pi^a(\vec{x},t)f_{ab}(\vec{x}-\vec{y})\pi^b(\vec{x},t)$ has a singularity at $\vec{k}=0$. For the existence of $\partial_{k_1}\cdots\partial_{k_n}f_{ab}(\vec{k})|_{\vec{k}=0}$, we need $|\int \mathrm{d}^dx f_{ab}(\vec{x})x_1\cdots x_n|<\infty$; thus, $f_{ab}(\vec{x})$ should behave as $x^{-r}$ ($r>n+d$) as $x\rightarrow\infty$.

If we explicitly break the $\text{U}(1)$ symmetry to open the bulk gap
by adding $-M_a^2\varphi_a^2/2g_a$ to the Lagrangian~\eqref{start}, the first term of Eq.~\eqref{effective} is replaced by 
\begin{equation}
\frac{m_1n_1\omega^2}{\sqrt{M_1^2+k_{\parallel}^2-\frac{\omega^2}{v_1^2}}}+\frac{m_2n_2\omega^2}{\sqrt{M_2^2+k_{\parallel}^2-\frac{\omega^2}{v_2^2}}}.
\end{equation}
Then the $k_{\parallel}$ dependence in the denominator becomes
subleading and the usual linear dispersion relation $\omega\propto k_{\parallel}$ is recovered.  In real space, the induced term is local $\propto\dot{u}^2(\vec{x},t)$ to leading order.

This understanding helps us to generalize our analysis. For example,
the domain wall of the $\mathbb{Z}_2$ symmetry-broken phase of the
real scalar $\phi^4$ theory should have the ordinary linear dispersion
relation since the bulk is gapped. Indeed, such domain wall fluctuation can be well described by the (local) Nambu-Goto action~\cite{Shifman}.  

Even when the bulk is gapped, if there are additional gapless degrees
of freedom on the domain wall, type B Nambu-Goldstone
bosons~\cite{WatanabeMurayama1} are possible, as recently discussed in
Ref.~\cite{Nitta}. However, they cannot have a fractional dispersion
relation unless one introduces a nonlocal term by hand.

Another essential ingredient in the above derivation is a finite
particle density.  That is, the same phenomenon cannot be realized by
relativistic superfluids at the zero chemical potential.  This can be
easily seen by setting $n_a=0$ (or equivalently $\mu_a=0$) in
Eqs.~\eqref{coupling} and \eqref{effective}. In the absence of
the $\omega^2/k_{\parallel}$ term in Eq.~\eqref{effective}, we add
the leading order term $O(\omega^2)$ to the Lagrangian and again find
a linear dispersion relation.  This basically means that, when the Lagrangian does not have
the $-n_a\dot{\varphi}_a$ term, the
domain wall fluctuation $u$ and the bulk phase fluctuation $\varphi_a$
are completely decoupled to quadratic order in the fluctuations.

As pointed out before by many authors~\cite{Mazets,Takeuchi,Kobyakov,Hayata}, the fluctuation of a
fluid surface with the dispersion relation $\omega\propto k^{3/2}$ is known to occur even in classical hydrodynamics~\cite{Landau}.  In this context, the mode is called a capillary wave.  In the classical fluid mechanics, the velocity field of an irrotational flow can be written as $\vec{v}=\vec{\nabla}\phi$, where $\phi$ is called the velocity potential. Further assuming the incompressibility and neglecting the dissipation and the gravitational potential, the pressure of a classical fluid can be expressed as~\cite{Landau}
\begin{equation}
\mathcal{P}=p_0-\rho\dot{\phi}-\frac{\rho}{2}(\vec{\nabla}\phi)^2+\cdots,\label{classical}
\end{equation}
where $\rho$ is the mass density of the fluid.  This should be compared to the pressure of the superfluid,
\begin{equation}
\mathcal{P}=\frac{\mu^2}{2g}-n\dot{\varphi}-\frac{n}{2m}(\vec{\nabla}\varphi)^2+\frac{n}{2m v^2}\dot{\varphi}^2+\cdots.
\end{equation}
We notice the formal correspondence $\varphi=m\phi$.  Therefore, the above derivation goes without changes for classical fluids, except that (i) we replace $\mathcal{P}$ in Eq.~\eqref{start} with the pressure Eq.~\eqref{classical} for a classical fluid and (ii) we take the infinite speed of sound limit $v\rightarrow\infty$ in, e.g., Eq.~\eqref{effective} to be consistent with the assumption of incompressibility.  For example, it is known that the potential on the surface $\phi|_{z=0}$ and the position of the surface $u$ are canonically conjugate to each other~\cite{Zakharov}.  This fact can be understood by Eq.~\eqref{coupling} and the correspondence $\phi=\varphi/m$.

Finally, let us comment on the effect of the gravitational potential.  Naively, it explicitly breaks the translational symmetry in the $z$ direction; hence, the NGB associated with the translational symmetry should open a gap.  Indeed, the effective Lagrangian may obtain a ``mass term" $-M^2 u^2/2$ with $M^2\equiv (m_1n_1-m_2n_2)g>0$ \footnote{Gravity also creates a density gradient.  However, its effect on the ``mass term" is higher order in $g$ and we ignore it during our discussions.}. However, it turns out that the domain wall fluctuation remains massless thanks to 
the interplay with bulk gapless modes.

Let us first discuss the incompressible limit $v_a\rightarrow\infty$.  Adding the mass term to the effective Lagrangian \eqref{effective}, one finds~\cite{Volovik,Pethick}
\begin{equation}
\omega=\sqrt{\frac{(m_1n_1-m_2n_2)gk_{\parallel}+\sigma k_{\parallel}^3}{m_1n_1+m_2n_2}}.\label{gravity}
\end{equation}
The dispersion is proportional to $k_{\parallel}^{1/2}$ in the long-wavelength limit. This mode is called the gravity wave of a fluid surface and exists in the incompressible classical fluid as well~\cite{Landau}.  For a finite $v_a$, one can no longer neglect $\omega^2/v_a^2$'s in the denominator of \eqref{effective} for a small $k_{\parallel}$ limit~\cite{Takeuchi}.  The dispersion in Eq.~\eqref{gravity} is valid only for the wave vector sufficiently bigger than
\begin{equation}
\max_{a=1,2}\frac{(m_1n_1-m_2n_2)g}{(m_1n_1+m_2n_2)v_a^2},
\end{equation}
but smaller than $\min_a 2\pi/\xi_a$, with $\xi_a=(2m_a\mu_a)^{-1}$ being the coherence length.

\subsection{Commutation relations}
Now let us come back to the consequence of Eq.~\eqref{coupling}. This
type of quadratic term linear in the time derivative usually implies
that conserved charges associated with these fields do not
commute~\cite{WatanabeMurayama1,WatanabeMurayama5}.

Indeed, from Noether's theorem, the $z$ component of the momentum operator $P_z$ is found to be
\begin{eqnarray}
P_z&=&\int\mathrm{d}^2x_{\parallel}\left[\int_u^\infty\mathrm{d}z\,j_1^0\partial_z\varphi_1+\int_{-\infty}^u\mathrm{d}z\,j_2^0\partial_z\varphi_2\right]\notag\\
&\simeq&\int\mathrm{d}^2x_{\parallel}(-n_1\varphi_1+n_2\varphi_2)|_{z=0}.
\end{eqnarray}
The formula for U(1) charges $Q_a$ in Eq.~\eqref{u1} can also be linearized as
\begin{equation}
Q_1\simeq-\int\mathrm{d}^2x_{\parallel}n_1u,\quad
Q_2\simeq\int\mathrm{d}^2x_{\parallel}n_2u.
\end{equation}
The canonical commutation relation mentioned below Eq.~\eqref{coupling} then suggests that the algebra of symmetry generators is centrally extended due to the presence of the domain wall:
\begin{eqnarray}
[P_z,Q_1]=iAn_1,\quad [P_z,Q_2]=-iAn_2,
\end{eqnarray}
where $A=\int\mathrm{d}^2x_{\parallel}$ is the total area of the
domain wall that diverges in the thermodynamic limit.  This result is
counterintuitive since $Q_a$ and $P_z$ naively commute.  Also, the
Jacobi identity among $J_x,P_y,Q_a$ ($\vec{J}$ is the angular
momentum) appears to prohibit such central extensions.  However, in
the presence of the domain wall, $J_x$ is no longer well defined, as it
changes the field configuration at the spatial boundary and we do not
have to impose the Jacobi identity (as discussed in Ref.~\cite{WatanabeMurayama4}).

The noncommuting algebra (central extensions) as a consequence of topological solitons or domain walls are recently discussed in Refs.~\cite{WatanabeMurayama4,Nitta}. Here we confirmed that the superfluid-superfluid interface of phase-separated two-component Bose-Einstein condensates is an example of such an extension.

Note that the counting rule of NGBs discussed in Refs.~\cite{WatanabeBrauner1,WatanabeMurayama1,Hidaka,WatanabeMurayama5} can be violated when the effective Lagrangian fails to be local because the proof assumes a local effective Lagrangian.

\section{Phonons in Wigner solid in $2+1$ dimensions}
The longitudinal phonon in a two-dimensional crystal sheet of
electrons (Wigner crystal) embedded in three-dimensional space is
completely analogous to the ripplon example.  It has the dispersion
relation $\omega\propto\sqrt{k_{\parallel}}$, due to the long-range Coulomb interaction~\cite{Giuliani}. Here we revisit this well-known example to clarify the similarity.

The long-range Coulomb interaction is mediated by photons propagating
in the three-dimensional space. To explicitly deal with this physics,
let us take the free photon Lagrangian plus the interaction to charges
in the system:
\begin{eqnarray}
-\int\mathrm{d}^3x\left[\frac{1}{16\pi}F_{\mu\nu}F^{\mu\nu}+A_\mu (j_e^\mu+j_p^\mu)\right]\label{FF}
\end{eqnarray}
where $F_{\mu\nu}=\partial_\mu A_\nu-\partial_\nu A_\mu$ and 
\begin{eqnarray}
j_e^\mu&=&n_0e\delta(z)(1-\nabla\cdot\vec{u},\dot{u}_x,\dot{u}_y,0),\\
j_p^\mu&=&-n_0e\delta(z)(1,0,0,0)
\end{eqnarray}
are the electron current including the effect of the lattice deformation and the background positive charge that neutralizes the total electric charge density at the equilibrium. 
Here we consider only the in-plane fluctuation $\vec{u}=(u_x,u_y)$.  

In the Lorentz gauge, the equation of motion of the gauge field can be solved as
\begin{eqnarray}
A^\mu(\vec{k}_{\parallel},k_z,\omega)=\frac{4\pi}{k_{\parallel}^2+k_z^2-\frac{\omega^2}{c^2}}(j_e^\mu+j_p^\mu)(\vec{k}_{\parallel},\omega).
\end{eqnarray}
Hence integrating out photons induces a term,
\begin{eqnarray}
&&-\frac{n_0^2e^2}{2}u^*_i\left[\int\frac{\mathrm{d}k_z}{2\pi}\frac{4\pi(k_\parallel^ik_\parallel^j-\delta^{ij}\frac{\omega^2}{c^2})}{k_{\parallel}^2+k_z^2-\frac{\omega^2}{c^2}}\right]u_j\notag\\
&&=-\frac{n_0^2e^2}{2}u_i^*\frac{2\pi(k_\parallel^ik_\parallel^j-\delta^{ij}\frac{\omega^2}{c^2})}{\sqrt{k_{\parallel}^2-\frac{\omega^2}{c^2}}}u_j.\label{Coulomb}
\end{eqnarray}
After taking the limit $c\rightarrow\infty$, this term reduces to the Coulomb interaction between excess charges $-n_0e\nabla\cdot\vec{u}$:
\begin{equation}
-\frac{1}{2}\int\mathrm{d}^2x\mathrm{d}^2y\vec{\nabla}_{\vec{x}}\cdot\vec{u}(\vec{x},t)\frac{n_0^2e^2}{|\vec{x}-\vec{y}|}\vec{\nabla}_{\vec{y}}\cdot\vec{u}(\vec{y},t)\label{coulomb2}
\end{equation}
in real space, which is again nonlocal.

To discuss the dispersion relation of the longitudinal mode $u_L(\vec{k}_{\parallel})\equiv\vec{k}_{\parallel}\cdot\vec{u}/k_{\parallel}$, we introduce the kinetic term $\int\mathrm{d}^2x(m n_0/2)\dot{\vec{u}}^2(\vec{x},t)$ to the Lagrangian \eqref{FF}.  Then, we find the well-known dispersion
 relation $\omega^2=2\pi n_0 k_{\parallel}/m$, as $k_{\parallel}$ in the denominator of Eq.~\eqref{Coulomb} reduces the one power of $k_{\parallel}$ in the numerator.  This derivation again clarifies the importance of the gapless nature of the surrounding modes.  When the difference of the dimensionality between the bulk and the crystal subspace is $m$, integrating out photons produces an interaction $\propto k_{\parallel}^{m-2}$ among the longitudinal component $k_{\parallel}u_L$, ending up with a $k_{\parallel}^{m/2}$ dispersion~\footnote{This evaluation is based on scaling and might miss logarithmic corrections.}.  Our analysis above corresponds to the case $m=1$.  The three-dimensional Wigner crystal, whose longitudinal phonon is gapped, corresponds to the case $m=0$.  Specifically, $m$ has to be an odd integer to get a fractional power from this mechanism.
 
Note that long-range interactions between spatial derivatives of
fields [e.g., Eq.~\eqref{coulomb2}] make dispersion relations
harder, while those among time
derivatives [e.g., Eq.~\eqref{effective}] have the opposite
effect. 

\section{Conclusion and Discussions}
In this paper, we clarified the origin of the fractional-power
dispersion relation of ripplons in a superfluid-superfluid interface.
Gapless modes in the bulk induce a nonlocal interaction, which takes
a nonanalytic form in Fourier space, leading to a
fractional-power dispersion relation of the domain wall fluctuation.

When there are gapless modes localized on the domain wall other than the ripple mode of the domain wall or the in-plane phonon modes, bulk gapless excitation can, in principle, induce a nonlocal coupling for them as well.

Softer modes in the bulk tend to have a more drastic effect on modes
localized on the domain wall.  It might be an interesting future work
to look for a realistic example of an interface of two bulk systems
which support NGBs with a softer dispersion relation.

The low-energy physics of a system can be qualitatively modified only
by coupling it to other gapless degrees of freedom.  When Goldstone
modes associated with spacetime symmetries are coupled to a Fermi
surface, for example, they get overdamped and can no longer be a good
particlelike excitation~\cite{Watanabe2014}.  This is again a result of the interaction with a continuous spectrum of electron-hole pair excitations.

We thank Tom\'a\v{s} Brauner for stimulating discussion and for critical reading of the draft, and Yoshimasa Hidaka, Hiromitsu Takeuchi, Muneto Nitta, and Michikazu Kobayashi for a useful discussion on the ripplon.  H. W. appreciates the financial support of the Honjo International Scholarship Foundation.  The work of H. M. was supported by the U.S. DOE under Contract No. DE-AC03-76SF00098, by the NSF under Grants No. PHY-1002399 and No. PHY-1316783, by JSPS Grant No. (C) 23540289, and by WPI, MEXT, Japan.

{\bf Note added in proof:}---After the competition of our work, we have been informed of several related studies in completely different contexts.  Reference~\cite{Rex} discussed a fractional-power spectrum in the entanglement Hamiltonian, which is a result of a nonlocal interaction generated by tracing out gapless modes.  Reference~\cite{Juan} obtained a NGB with a fractional-power dispersion relation in the context of holography.

\bibliography{references}
\end{document}